\documentclass[prd,aps,preprint,preprintnumbers,nofootinbib]{revtex4}
\usepackage{epsfig,footnote}
\usepackage{ulem}
\usepackage{color}
\usepackage{array}
\usepackage{amssymb}
\usepackage{amsmath}
\usepackage{physics}
\usepackage{graphicx,subfigure}
\usepackage{booktabs}  
\usepackage{array}
\usepackage{longtable}
\usepackage[flushleft]{threeparttable}
\usepackage{verbatim}
\usepackage{amsfonts}
\usepackage{hyperref}
\usepackage{cancel}

\begin{document}
\title{Precise theoretical prediction on branching fractions and polarizations of $D \to V V$ decays }
\author{Jing Ou-Yang$^{1}$, Hui Zheng$^{1}$, Run-Hui Li$^{1}$ and 
Si-Hong Zhou$^{1,2}$\footnote{corresponding author: shzhou@imu.edu.cn}}

\affiliation{$^{1}$Inner Mongolia Key Laboratory of Microscale Physics and Atom Manufacturing, 
School of Physical Science and Technology, 
Inner Mongolia University, Hohhot 010021, China\\
$^{2}$Center for Quantum Physics and Technologies, 
Inner Mongolia University, Hohhot 010021, China}
\date{\today}

\begin{abstract}
We present a precise and systematic analysis of 
$D \to V V$ decays within the factorization-assisted topological-amplitude 
approach, where $D$ denotes the set $\{D^0, \, D^+,\, D^+_s\}$ and $V$ 
represents the vector mesons $\rho, K^*, \omega$, and $\phi$. Given the 
limited current experimental data, the factorization-assisted topological-amplitude
approach serves as a available phenomenological 
framework for predicting charmed meson decays to both vector mesons.
In this framework, incorporating flavor SU(3) symmetry-breaking effects,
we can express nonfactorizable contributions of different modes as a 
minimal set of universal parameters globally fitted to experimental data. 
Utilizing 36 experimental data points for $D \to VV$ decays, we precisely 
extract ten nonfactorizable parameters associated with the $C$ and $E$ 
topological diagrams with $\chi^2/\mathrm{d.o.f.}=8.43$. We find that a 
large strong phase in the longitude $E$ amplitude cause strong destructive 
interference with the $C$ longitudinal component, yielding $f_\parallel >f_L $, 
contrary to the naive factorization predictions. Additionally, for modes 
processing exclusively by the $E$ diagram, the amplitude hierarchy 
$|S|<|D|$ leads to a $D$-wave branching fraction larger than that of the 
$S$ wave. This explains recent observations that contradict $S$-wave 
dominance predictions. The predicted branching fractions and polarizations 
for 28 decay modes are consistent with existing experimental data.
 Unobserved modes, especially those with branching fractions of the order 
 $10^{-3}\sim10^{-2}$, the $D$-wave-dominated modes, and modes exhibiting
  $f_\parallel >f_L $, await measurement by BESIII, STCF, Belle II, and LHCb.
  \end{abstract}

\maketitle

\section{Introduction}\label{sec:1}

Since 2003, BaBar~\cite{BaBar:2003zor} and Belle~\cite{Belle:2003ike}
have measured large transverse polarization fractions ($\sim 50\%$) in 
$B \to \phi K^*$, respectively, contradicting the fraction expectation of 
dominant longitude polarization fractions, $f_L=1-\mathcal{O} (m_V^2
/m_B^2)$. Similar anomalies were observed in other penguin-dominated 
strangeness-changing decays, such as 
$B^+ \to \omega K^{*+}$, $B^+ \to \rho^0\, K^{*+}$, and $B_s \to \phi \phi$, 
sparking extensive theoretical interest. Although transverse polarization 
fractions in the charm sector are not as heavily suppressed as in $B$ 
decays, they still significantly 
deviate from naive factorization estimates. For instance, Mark III 
Collaboration reported a transverse branching fraction of 
$\mathcal{B} (D^0 \to \bar{K}^{*0} \rho^0)_\mathrm{T}=(1.6 \pm 0.6)\%$,
comparable to the total branching fraction of
$\mathcal{B} (D^0 \to \bar{K}^{*0}\rho^0)_\mathrm{tot}=(1.59 \pm 0.35)\%$~\cite{MARK-III:1991fvi}, 
and partial wave analysis yields $S$-, $P$- and 
$D$-wave components of 
 $(3.1 \pm 0.6)\%,\, <3 \times 10^{-3}$ and $(2.1 \pm 0.6)\%$, respectively.
This hierarchy contradicts the naive prediction $|S|^2>|P|^2>|D|^2$. 
Subsequently, BESIII found $D$-wave dominance in $D^0 \to K^{*-} \rho^+, \, 
\bar{K}^{*0} \rho^0$, $\rho^+\rho^-$ and $P$-wave dominance in 
$D_s^+ \to \bar{K}^{*0} \rho^+, \, K^{*-} \rho^0$~\cite{BESIII:2017jyh,
BESIII:2023exz,BESIII:2022bvv}. Additionally, the $D$-wave dominance  
in $D^0 \to \rho^0 \rho^0$ was reported by FOCUS~\cite{FOCUS:2007ern}, 
with partial-wave analyses further done in Ref.~\cite{dArgent:2017gzv}.

To understand these ``polarization anomalies" in $D \to VV$ decays 
within QCD remains challenging due to the intermediate charm mass 
scale, $m_c$. 
Unlike $B$ decays, where heavy quark expansion ($1/m_b$) enables 
successful applications of QCD factorization, perturbative QCD, and 
soft-collinear effective theory, the mass $m_c$ is not sufficiently large 
to allow for such an expansion. Early studies of $D \to VV$ decays 
relied on naive factorization estimates~\cite{ Kamal:1990ky,
Bauer:1986bm,Cheng:2010rv,Uppal:1992se}, pole-dominance 
model~\cite{Bedaque:1993fb}, heavy quark effective Lagrangian 
combined with chiral perturbative theory~\cite{Bajc:1997ey}, or 
model-independent symmetry based approaches, including flavor 
SU(3) symmetry~\cite{Kamal:1990ky} and broken flavor SU(3) 
symmetry models~\cite{Hinchliffe:1995hz}. These studies typically 
focus on only a few specific modes. Recent works have extended 
this to systematic analyses of all Cabibbo-favored and singly 
Cabibbo-suppressed $D^0 \to VV$ decays using quark model with 
final state interactions (FSIs)~\cite{Cao:2023csx}, and all $D^0\to VV$ 
decays have been studied in naive factorization~\cite{Cheng:2024hdo}.
While abundance data for $D \to PP$ and $D \to PV$ decays have 
enabled systematic phenomenological studies via topological diagram 
approach~\cite{Cheng:2021yrn,Cheng:2024hdo}, the 
factorization-assisted topological-amplitude (FAT) 
approach~\cite{Li:2013xsa,Zheng:2025ryf}, and quark 
model~\cite{Wang:2026ghd}, $D \to VV$ decays are experimentally 
more challenging. Consequently, the resulting scarcity and limited 
precision of data have prevented their study within the topological 
diagram approach. Leveraging current available data, the FAT 
approach offers a viable framework to predict  all $D_{(s)} \to VV$ 
mode, encompassing Cabibbo-favored, singly Cabibbo-suppressed, 
and doubly Cabibbo-suppressed decays.

The FAT approach builds upon the conventional topological diagram 
approach~\cite{Cheng:2010ry,Cheng:2012wr}. In the topological 
diagram approach, decay amplitudes are categorized into distinct 
topological diagrams according electroweak interactions. As the 
weak interaction is factorized automatically from strong interaction 
due to their distant scales, the QCD effects including perturbative 
and nonperturbative contributions (including FSIs), are effectively 
encapsulated within these topological amplitudes. Consequently, 
when these amplitudes (excluding the common factor $G_F/\sqrt {2}\, 
V_\mathrm{CKM}$) are extracted directly from experimental data,
they inherently contain all QCD effects. However,  the conventional 
topological diagram approach relies on flavor SU(3) symmetry to 
reduce the number of free parameters so as to improve fit quality. 
It is well established that the SU(3)-breaking effects can reach 
$20-30\%$ in $B$ decays~\cite{Zhou:2015jba, Zhou:2016jkv} 
and are expected to be even larger in $D$ decays. This significant 
symmetry breaking limits the prediction power of the conventional 
diagrammatic approach. The FAT approach addresses this limitation 
by incorporating SU(3) breaking effects assisted by factorization. 
Especially, we factor out the form factors and decay constants from 
the topological diagram amplitudes. As a result, the remaining 
nonfactorizable contributions become universal across all decay 
modes and can be characterized by a minimal set of free parameters,
determined through a global fit to all available experimental data. 

The FAT framework was originally developed for 
charm meson decays~\cite{Li:2012cfa, Li:2013xsa,Zheng:2025ryf},
and later extended to $B$-meson decays 
by one of us (S.-H. Z.) and 
collaborators~\cite{ Zhou:2015jba, Zhou:2016jkv, Wang:2017hxe} 
(see ~\cite{Qin:2021tve} for a review). 
It has been applied to $D^0-\bar D^0$ mixing~\cite{Jiang:2017zwr}, 
 $K_S^0-K_L^0$ asymmetries~\cite{Wang:2017ksn},
{\it CP} violation in charm decays~\cite{Yu:2017oky}, and
the extraction of the Cabibbo-Kobayashi-Maskawa (CKM)
phase $\gamma$ from charmless two-body $B$ 
decays~\cite{Zhou:2019crd}.
Recent extensions include quasi-two-body $B$ 
decays~\cite{Zhou:2021yys,Zhou:2023lbc,Zhou:2024qmm,Ou-Yang:2025ije} 
and $D$ decays~\cite{Zhou:2025nao,Wang:2025rkr}. 
Notably, the FAT approach has successfully 
explained the ``polarization anomalies"
in $B \to VV$ decays~\cite{Wang:2017hxe} by 
predicting reduced branching fractions and longitudinal 
polarization fractions for color-suppressed decays and
attributing large transverse polarization fractions observed 
in the penguin-dominant modes to a single 
transverse amplitude.
In this work, we apply the the FAT approach to
systematically analyze $D \to VV$ decays.
We aim to resolve the polarization anomalies in the charm sector 
and provide robust predictions as references for 
upcoming experimental measurements.

The remainder of this paper is organized as follows. 
 In Sec.~\ref{sec:2}, we introduce the theoretical framework. 
Numerical results and detailed discussions are presented in Sec.~\ref{sec:3}. 
Finally, our conclusions are summarized in Sec.~\ref{sec:4}.
 

\section{Factorization Amplitudes for Topological Diagrams}\label{sec:2}
In this section, we first introduce the decay amplitudes for $D \to VV$ 
decays in various bases, along with their expressions within naive 
factorization. We then proceed to analyze the three polarization 
amplitudes under the FAT framework.
\subsection{Decay amplitude decomposition and naive factorization}
For a $D$ meson with four-momentum $p_D$ decaying into two vector 
mesons $V_1(m_1, p_1, \eta^*)$ and $V_2(m_2, p_2, \epsilon^*)$ with 
polarization vectors $\eta^*$ and $\epsilon^*$, the decay amplitude, 
based on Lorentz decomposition, reads
\begin{equation}
\mathcal{A}_{D\to V_1V_2} = i\eta^{*\mu}\, \epsilon^{*\nu}\, \left(g_{\mu\nu}\, S_1\,  -\,  
\frac{p_{D\mu}p_{D\nu}}{m_D^2}\, S_2 \, +\, 
 i\epsilon_{\mu\nu\rho\sigma}\, \frac{p_1^\rho p_2^\sigma}{p_1\cdot p_2}\, S_3\right)\, .
\label{eq:amplitude_decomposition}
\end{equation}
The amplitudes with definite helicity ($h=0,\, +,\, - $) are expressed as
\begin{align}\label{eq:Apm}
\begin{aligned}
\mathcal{A}^0 &= \mathcal{A}\left[D \to V_1(p_1, \eta_0^*)\, V_2(p_2, \epsilon_0^*)\right]
= i\frac{m_D^2}{2m_1m_2}\, \left(S_1 \, -\,  \frac{S_2}{2}\right)\, , \\
\mathcal{A}^{\pm} &= \mathcal{A}\left[D \to V_1(p_1, \eta_{\pm}^*)\, V_2(p_2, \epsilon_{\pm}^*)\right]
= i(S_1 \, \mp\,  S_3)\, .
\end{aligned}
\end{align}
Within the naive factorization framework, the helicity amplitudes can be 
formally expressed as
\begin{equation}
\mathcal{A}^{h} = \frac{G_F}{\sqrt{2}}V_{CKM}\bra{V_1^{h}}(\bar{c}q)_{V-A}\ket{D}
\bra{V_2^{h}}(\bar{q}q')_{V}\ket{0}\, ,
\label{eq:factorization}
\end{equation}
By neglecting $m_i^2$ contributions, the three helicity amplitudes 
are explicitly given by
\begin{align}\label{eq:Aminus_simple}
\begin{aligned}
\mathcal{A}^0 &= i\frac{G_F}{\sqrt{2}}\, V_{CKM}\, f_2\, m_D^2\, A_0^{DV_1}(m_2^2)\, ,  \\
\mathcal{A}^{+} &= i\frac{G_F}{\sqrt{2}}\, V_{CKM}\, f_2\, m_2\, 
\left\{-(m_D+m_1)\, A_1^{DV_1}(m_2^2)\, +\, (m_D-m_1)\, V^{DV_1}(m_2^2)\right\}\, ,  \\
\mathcal{A}^{-} &= i\frac{G_F}{\sqrt{2}}\, V_{CKM}\, f_2\, m_2\, 
\left\{-(m_D+m_1)\, A_1^{DV_1}(m_2^2)\, -\, (m_D-m_1)\, V^{DV_1}(m_2^2)\right\}\, ,
\end{aligned}
\end{align}
where $A_0^{DV_1}(m_2^2)$, $A_1^{DV_1}(m_2^2)$ and $V^{DV_1}(m_2^2)$ 
denote the transition form factors, as defined in Refs.~\cite{Wirbel:1985ji,Bauer:1986bm},
and $f_2$ is the decay constant of the emitted vector meson.
Given the  $m_2/m_D$ suppression of $\mathcal{A}^\pm$ relative to the $\mathcal{A}^0$
and the $A_1$ and $V$ cancellation, the amplitudes follow the hierarchy
\begin{equation}
\mathcal{A}^0 : \mathcal{A}^- : \mathcal{A}^+ =
 1 : \frac{\Lambda_\mathrm{QCD}}{m_c} : \left( \frac{\Lambda_\mathrm{QCD}}{m_c} \right)^2\, .
\end{equation}

In general, decay amplitudes admit several equivalent representations. 
Especially, the helicity amplitudes $\mathcal{A}^{0,+,-}$ relate to the spin 
amplitudes in the transversity basis ($\mathcal{A}^{L,\parallel, \perp,}$) 
as follows:
\begin{align}
\mathcal{A}_L &= \mathcal{A}^0 \, ; \quad 
\mathcal{A}_\parallel = \frac{\mathcal{A}^+ + \mathcal{A}^-}{\sqrt{2}}\, ; \quad 
\mathcal{A}_\perp = \frac{\mathcal{A}^+ - \mathcal{A}^-}{\sqrt{2}} \, .
\end{align}
In the transversity basis, the three amplitudes can be further simplified as
\begin{align} \label{NFres}
\begin{aligned}
\mathcal{A}_L &= i \frac{G_F}{\sqrt{2}} \, V_{CKM}\, f_2 \, m_D^2 \, A_0^{DV_1}(m_2^2)\, ,
 \\
\mathcal{A}_\parallel &= -i G_F\,  V_{CKM}\, f_2 \, m_2\,  (m_D + m_1) \, A_1^{DV_1}(m_2^2)\, ,
 \\
\mathcal{A}_\perp &= i G_F \, V_{CKM}\, f_2 \, m_2 \, (m_D - m_1)\,  V^{DV_1}(m_2^2)\, ,
\end{aligned}
\end{align}
As experimental data for $D \to VV$ decays are predominantly 
presented in terms of partial-wave  ($S, P, D$) branching fractions, 
it is necessary to convert the transversity amplitudes into partial-wave amplitudes via
\begin{align}\label{SPDdef}
S &= \frac{1}{\sqrt{3}}(-\mathcal{A}_L + \sqrt{2}\mathcal{A}_{\parallel})  \, ; \quad 
P =\mathcal{A}_{\perp}\, ; \quad 
D=  \frac{1}{\sqrt{3}}(\sqrt{2}\mathcal{A}_L +\mathcal{A}_{\parallel}) \, .
\end{align}
Equations.(\ref{NFres}) and (\ref{SPDdef}) reveal that, in the naive factorization approach,
the polarization amplitude hierarchies satisfy
 $|\mathcal{A}_\parallel|^2 \geq |\mathcal{A}_L|^2>|\mathcal{A}_\perp|^2$,
and $|S|^2 \geq |P|^2>|D|^2$~\cite{Cheng:2024hdo}.  

\subsection{Polarization amplitudes within the FAT framework}
\begin{figure} [htb]
\begin{center}
\scalebox{1}{\epsfig{file=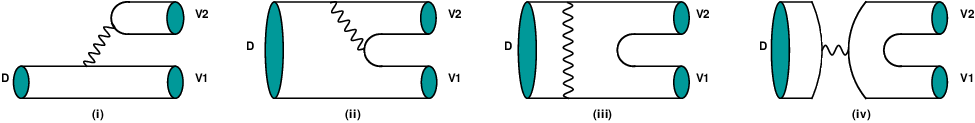}}
\caption{Topological diagrams for $D \to V_1 V_2 $ with the wave line representing a W boson: 
(i) the color-favored emission diagram T, 
(ii) the color-suppressed emission diagram C, 
(iii) the W-exchange diagram E, 
and (iv) W-annihilation diagram A.}
\label{tree}
\end{center}
\end{figure}

The $D \to V V$ decay is a weak process induced primarily by the tree-level 
quark transition $c \to d(s) u \bar{d} (\bar s)$.
Although penguin contributions ($c \to \,u  q \bar{q} \, \text{with}\, q=u, d, s)$)
exist, they are neglected in branching fraction calculations due to
suppression by both small Wilson coefficients and CKM matrix elements, 
Based on the weak interaction, the tree-level topological diagrams
 for $D  \to  V_1 V_2$ decays are conventionally classified into four categories:
(i) color-favored emission diagram $T$, 
(ii) color-suppressed emission diagram $C$,  
(iii) $W$-exchange diagram $E$, and 
(iv) $W$-annihilation diagram $A$, which are 
as illustrated in Fig.~\ref{tree}. 
Here, we factorize the weak interaction from strong interaction, 
thereby incorporating all perturbative and nonperturbative QCD 
corrections into the topological diagrams.
While the conventional topological diagram approach treats these four topological 
amplitudes as unknown parameters fitted under SU(3) symmetry,
the FAT approach explicitly accounts for SU(3)-breaking effects 
via factorization. In the following, we analyze the four
topological amplitudes ($T$, $C$, $E$, and $A$) within the FAT framework.

Firstly, the $T$ diagram is typically treated as factorizable. 
To minimize the number of free parameters, 
we directly adopt the naive factorization results, analogous to
Eq.(\ref{NFres}), including the relevant Wilson coefficients.
The $T$ amplitudes for the three polarization states are expressed as
\begin{align}  \label{eq:T}
\begin{aligned} 
    T^0 &= i \frac{G_F}{\sqrt{2}}\,  V_\mathrm{CKM} \, a_1 (\mu)\, m_D^2\,  f_2 \,  A_0^{DV_1}(m_2^2)\, ,\\
    T^{\parallel} &= -i G_F\,  V_\mathrm{CKM}\,  a_1 (\mu) \,m_2\,  (m_D + m_1) \,  f_2 \, A_1^{DV_1}(m_2^2)\, ,\\
    T^{\perp} &= i G_F\,  V_\mathrm{CKM} \, a_1 (\mu)\, m_2 \, (m_D - m_1) \,  f_2 \,  V^{DV_1}(m_2^2)\, ,
\end{aligned}
\end{align}
where $a_1 (\mu) = C_2 (\mu) + C_1 (\mu)/3$ is the effective 
Wilson coefficient for four-quark operators. The scale parameter 
$\mu$, representing the energy release,
is treated as a single free parameter constrained to be below $m_c/2$. 
In contrast to the conventional diagrammatic approach, 
where describing the three decay amplitudes requires at least five free parameters 
(three magnitudes and two strong phase for $T$ amplitudes)
to be fitted from experimental data, our formulation introduces only one free parameter,
$\mu$, for all $T$ amplitudes. 
Furthermore, SU(3)-breaking effects are naturally incorporated
through the use of distinct decay constants and form factors for different decay modes.

The remaining three diagrams ($C, E, $ and $A$) are dominated by  
nonfactorizable contributions. 
For the $C$ diagram, after factorizing out form factor and 
decay constant to account for SU(3)-breaking effects, the residual 
contribution is expected to be universal across different processes.
We introduce two sets of unknown parameters, 
$\chi_C^{0\, (\parallel,\, \perp)}$ and $\phi_C^{0\, (\parallel,\, \perp)}$, 
representing the magnitudes and associate strong phases for each
 polarization amplitude, respectively. These amplitudes 
are expressed as
\begin{align}  \label{eq:C}
\begin{aligned} 
  C^0 &= i \frac{G_F}{\sqrt{2}}\,  V_\mathrm{CKM} \, \chi_C^0\,  \mathrm{e}^{i \phi_C^0}\, m_D^2\,   f_2 \, A_0^{DV_1}(m_2^2)\, , \\
  C^{\parallel} &= -i G_F\,  V_\mathrm{CKM} \, \chi_C^{\parallel} \, \mathrm{e}^{i \phi_C^{\parallel}} \, 
  m_2\,  (m_D + m_1) \, \, f_2\,   A_1^{DV_1}(m_2^2)\, , \\
  C^{\perp} &= i G_F \, V_\mathrm{CKM} \, \chi_C^{\perp}\, \mathrm{e}^{i \phi_C^{\perp}}\, m_2 \, (m_D - m_1)\,  f_2 \,  V^{DV_1}(m_2^2)\, .
\end{aligned}
\end{align}

Similarly, for the $E$ diagram, 
we factorize the relevant meson decay constants to characterize
SU(3)-breaking effects. The polarized amplitudes are given by
\begin{align}  \label{eq:E}
\begin{aligned} 
E^0 &= i \frac{G_F}{\sqrt{2}} \, V_\mathrm{CKM}\, \chi_E^0 \, \mathrm{e}^{i \phi_E^0}\, m_D^2 \,
 f_D\,  f_1 \, f_2 \,  \frac{1}{f_{\rho}^2}\, , \\
E^{\parallel} &= -i G_F \, V_\mathrm{CKM}\,  \chi_E^{\parallel} \, \mathrm{e}^{i \phi_E^{\parallel}} \,
m_D\, (m_1 +m_2)\, f_D\,  f_1 \, f_2 \,  \frac{1}{f_{\rho}^2}\, , \\
E^\perp &= i G_F\, V_\mathrm{CKM}\, \chi_E^\perp\,  \mathrm{e}^{i \phi_E^\perp} \, m_D^2\,  
 f_D\,  f_1\,  f_2 \, \frac{1}{f_{\rho}^2}\, ,
\end{aligned}
\end{align}
where the parameters $\chi_E^{0,\, \parallel,\, \perp }$ 
are dimensionless quantities normalized to $f_{\rho}^2$. 

The $A$ diagram amplitudes follow a parametrization
similar to the $E$ diagram, substituting the parameters
$\chi_E^{0,\, \parallel,\, \perp }$ and $\phi_E^{0,\, \parallel,\, \perp }$
with $\chi_A^{0,\, \parallel,\, \perp }$ and  $\phi_A^{0,\, \parallel,\, \perp }$:
\begin{align}  \label{eq:A}
\begin{aligned} 
A^0 &= i \frac{G_F}{\sqrt{2}} \, V_\mathrm{CKM}\, \chi_A^0\,  \mathrm{e}^{i \phi_A^0}\,
m_D^2 \,  f_D \, f_1 \, f_2\,  \frac{1}{f_{\rho}^2}\, , \\
A^{\parallel} &= -i G_F\,  V_\mathrm{CKM} \,  \chi_A^{\parallel} \, \mathrm{e}^{i \phi_A^{\parallel}} \, 
m_D\, (m_1 +m_2)\,  f_D \, f_1 \, f_2 \,  \frac{1}{f_{\rho}^2}, \\
A^\perp &= i G_F \, V_\mathrm{CKM}\,  \chi_A^\perp \, \mathrm{e}^{i \phi_A^\perp} \,  
m_D^2\,  f_D\,  f_1\,  f_2\,  \frac{1}{f_{\rho}^2}.
\end{aligned}
\end{align}
Nevertheless, we neglect the $A$ diagram in later analysis because its contribution 
is negligible.
Attempting to include these parameters in 
our fitting program, 
we fail to obtain stable solutions due to the limited precision of current experimental data.

In total, we have nine magnitudes and nine strong phases, along with the scale parameter $\mu$,
to be fitted simultaneously to the the experimental data. 
With these fitted parameters, we can predict the branching fractions and two out of the three
polarization fractions ($f_L,\, f_\parallel,\, f_\perp$)
for all $D_{\bar{q}} \to VV (\bar{q} = \bar{u}, \bar{d}, \bar{s})$ based on the standard definitions
as follows
 \begin{align}
 \begin{aligned} 
\Gamma &\equiv \frac{|\mathbf{p}|}{8 \, \pi\,  m_D^2} 
\frac{|\mathcal{A}_L|^2 \, + \, |\mathcal{A}_\parallel|^2 \, +\, 
 |\mathcal{A}_\perp|^2 \, + \, |\overline{\mathcal{A}}_L|^2 \, +\, 
  |\overline{\mathcal{A}}_\parallel|^2 \, +\,  |\overline{\mathcal{A}}_\perp|^2}{2}\, ,\\
  &= \frac{|\mathbf{p}|}{8 \, \pi\,  m_D^2} 
\frac{|S|^2 \, + \, |P|^2 \, +\,  |D|^2 \, + \, |\overline{S}|^2 \, +\, 
  |\overline{P}|^2 \, +\,  |\overline{D}|^2}{2}\, ,
  \end{aligned}
\end{align}
and 
 \begin{align}
f_{L,\parallel,\perp}^D=\frac{\Gamma_{L,\parallel,\perp}}{\Gamma} 
= \frac{|\mathcal{A}_{L,\parallel,\perp}|^2}{|\mathcal{A}_L|^2 + |\mathcal{A}_\parallel|^2 + |\mathcal{A}_\perp|^2}\, ,
\end{align}
respectively.

\section{Numerical Results and discussions}\label{sec:3}
\subsection{Input parameters}
The input parameters are categorized into
(i) CKM matrix elements and Wilson coefficients;
(ii) meson masses and decay constants;
and (iii) $D_{(s)} \to V$ transition form factors. 
\begin{table}[!h]
\centering
\caption{\label{masses}Masses $m_V$ and decay constant of mesons (in units of MeV). }
\begin{tabular}{cccccccccccccc}
\hline
Meson & Mass & Decay constant  \\
\hline
$D^{\pm }$ & 1869.66 & $212.0\pm 0.7$ \\
$D^{0}$ & 1864.84 & $212.0\pm 0.7$ \\
$D_{s}$ & 1968.35 & $249.9\pm 0.5$ \\
$\rho$ & 775.26 & $223.7 \pm 11.2$\\
$\omega$ & 782.66 & $ 182.4\pm 9.1$\\
$\phi$ & 1019.46 & $236.3 \pm 11.8$\\
$K^{*+}$ & 891.67 & $214.7 \pm 10.7$\\
$K^{*0}$ & 895.55 & $214.7 \pm 10.7$\\
\hline
\end{tabular}
\end{table}
The CKM matrix elements are taken from the 
PDG~\cite{ParticleDataGroup:2024cfk},
and the Wilson coefficients, $C_{1(2)}(\mu)$, for 
$D$-meson decays follow Eqs.~(B1) and  (B2) in 
the Appendix in Ref.~\cite{Li:2012cfa}. The
masses and decay constants of the $D$ mesons 
and vector mesons are listed in Table.~\ref {masses}. 
Specifically, all masses and $D_{(s)}$ meson decay 
constants are provided by the 
PDG~\cite{ParticleDataGroup:2024cfk}. The decay 
constants of vector mesons ($\rho, \, \omega,\, \phi,\, K^*$)
have not been measured experimentally but are calculated 
in several theoretical approaches, such as quark model~\cite{Verma:2011yw},
covariant light front approach~\cite{Cheng:2003sm}, 
light-cone sum rules~\cite{Ball:2006eu,Bharucha:2015bzk}, 
QCD sum rules~\cite{Gelhausen:2013wia}, etc. Given the variations 
among these theoretical results, we adopt the numerical values shown 
in Table.~\ref {masses} and keep a $5\%$ uncertainty to them.

The transition form factors of $D \to V$ are also unmeasured.
We adopt values from Ref.~\cite{Bauer:1986bm}, calculated  
at zero recoil momentum ($q^2=0$) in Ref.~\cite{Wirbel:1985ji} 
using relativistic oscillator wave functions. The corresponding 
values at $q^2=0$ are presented in Table.~\ref{tab:ff} and a 
$10\%$ uncertaintiy is assigned to these form factors. 
Their $q^2$ dependence is parametrized using the nearest 
pole dominance form~\cite{Wirbel:1985ji,Bauer:1986bm} as 
follows
\begin{equation}\label{eq:ffdipole}
F_i (q^2)=\frac{F_i (0)}{1- q^2/m^2_{\text{pole}}}\, ,
\end{equation}
where $F_{i}$ represents form factors $A_{0}$, $A_{1}$ or $V$, 
and $m_{\rm pole}$ is the mass of the corresponding pole state, 
which are displayed in Table~\ref{poles} from Refs.~\cite{Wirbel:1985ji,Bauer:1986bm}. 
Specifically, we use 
$m(0^-,\bar{d}c)$ for $A_{0}^{D \rho, D \omega, D_s K^*}$
and $m(0^-,\bar{s}c)$ for $A_{0}^{D K^*, D_s \phi}$; 
$m(1^+,\bar{d}c)$ for $A_{1}^{D \rho, D \omega, D_s K^*}$ 
and $m(1^+,\bar{s}c)$ for $A_{1}^{D K^*, D_s \phi}$;
$m(1^-,\bar{d}c)$ for $V^{D \rho, D \omega, D_s K^*}$;
and $m(1^-,\bar{s}c)$ for $V^{D K^*, D_s \phi}$. 
\begin{table} [htbp]
\caption{Form factors at zero recoil momentum for $D \to V$ transitions~\cite{Bauer:1986bm}. }\label{tab:ff}
\vspace{2mm}
\centering
\begin{tabular}{|c|c|c|c|c|c|c|c|c|c|c|}
\hline& \quad
$~~V^{D\to \rho}~~$ \quad&
$~~V^{D\to K^*}~~$ & \quad
$~~V^{D_s\to K^*}~~$ & \quad
$~~V^{D\to \omega}~~$& \quad
$~~V^{D_s\to \phi}~~$\\
\hline
$~V(0)~$&
1.225&
1.226&
1.250 &
1.236&
1.319\\
\hline&
$~~A_{0}^{D\to \rho}~~$&
$~~A_{0}^{D\to K^*}~~$ &
$~~A_{0}^{D_s\to K^*}~~$ &
$~~A_{0}^{D\to \omega}~~$&
$~~A_{0}^{D_s \to \phi}~~$ \\
\hline
$~A_0(0)~$&
0.669&
0.733&
0.634&
0.669&
0.700  \\
\hline&
$~~A_{1}^{D\to \rho}~~$&
$~~A_{1}^{D\to K^*}~~$ &
$~~A_{1}^{D_s\to K^*}~~$ &
$~~A_{1}^{D\to \omega}~~$&
$~~A_{1}^{D_s \to \phi}~~$ \\
\hline
$~A_1(0)~$&
0.775&
0.880&
0.717&
0.772&
0.820  \\
\hline
\end{tabular}
\end{table}

\begin{table}[htbp]
\centering
\caption{\label{poles}Values of pole masses used in Eq.(\ref{eq:ffdipole}) (in units of GeV).}
\vspace{0.2 cm}
\begin{tabular}{ccccc}
\hline
Current \quad &  \quad$m(0^-)$  \quad &  \quad$m(1^-)$ \quad  & \quad $m(0^+)$ \quad  &  \quad$m(1^+)$  \\
\hline
$\bar{d}c$ & 1.87 & 2.01 & 2.47 & 2.42 \\
$\bar{s} c$ & 1.97 & 2.11 & 2.60 & 2.53 \\
\hline
\end{tabular}
\end{table}

\subsection{The nonfactorizable parameters}
The free parameters in the topological diagram amplitudes 
defined in Eqs.~(\ref{eq:T})-(\ref{eq:A}) consist of $19$ parameters: 
$\chi_C^{0,\parallel,\perp}$, $\chi_E^{0,\parallel,\perp}$, and 
$\chi_A^{0,\parallel,\perp}$, along with their associated strong phases 
$\phi_E^{0,\parallel,\perp}$, $\phi_C^{0,\parallel,\perp}$, and 
$\phi_A^{0,\parallel,\perp}$, as well as the factorization scale $\mu$ 
[appearing in the Wilson coefficients $a_{1}(\mu)$ of the $T$ diagram 
amplitude in Eq.(\ref{eq:T})]. In the experimental sector, the total and 
partial branching fractions of $D \to VV$ have been measured by 
BESIII~\cite{BESIII:2017jyh,BESIII:2021dot,BESIII:2021qfo,
BESIII:2021raf,BESIII:2022bvv,BESIII:2023exz,BESIII:2023qgj}, 
LHCb~\cite{LHCb:2017swu,LHCb:2018mzv}, CLEO-c~\cite{dArgent:2017gzv}, 
FOCUS~\cite{FOCUS:2007ern}, and Mark III Collaborations~\cite{MARK-III:1991fvi}, 
where all measurements of $D^+$ and $D_s^+$ decays are provided by BESIII. 
The data used in the global fit are listed in Tables~\ref{CF}-\ref{SCS2}, 
taken from the PDG~\cite{ParticleDataGroup:2024cfk} and recent BESIII measurements.
In total, 36 data points are available to constrain the 19 free parameters.
However, we observe that the number of free parameters is too large 
to be precisely determined by the limited experimental data, thereby 
diminishing the predictive power of the FAT approach.

Given the very small values of $\chi_E^\perp$ found in preliminary fit, 
we neglect the parameter $\chi_E^\perp$ and its associated strong phase 
$\phi_E^\perp$. Moreover, as the $A$ diagram amplitude is typically smaller 
than that of the $E$ diagram, we ignore the $A$ diagram contributions entirely.
Consequently, by setting
 $\chi_E^\perp = \chi_A^0 = \chi_A^\parallel= \chi_A^\perp = 0$ and
 $\phi_E^\perp = \phi_A^0= \phi_A^\parallel= \phi_A^\perp = 0$,
 the number of free parameters is reduced to 11. These remaining parameters 
 are subsequently determined through a global fit to 36 experimental data points. 
The best-fitted values, together with their corresponding uncertainties, are list as 
follows,
\begin{align}\label{parameter}
\begin{aligned}
\mu=465.14 \pm 0.12\, \, \, &\text{MeV},\\
\chi_C^{0}=0.55 \pm 0.03,~~~&\phi_C^{0}=0.87 \pm 0.10,\\
\chi_C^{\parallel}=0.45 \pm 0.03,~~~&\phi_C^{\parallel}=0.77 \pm 0.11, \\
\chi_C^{\perp}=0.37 \pm 0.02,~~~&\phi_C^{\perp}=0.36 \pm 0.04, \\
\chi_E^{0}=0.24 \pm 0.07,~~~&\phi_E^{0}=2.23 \pm 0.30, \\
\chi_E^{\parallel}=0.25 \pm 0.03,~~~&\phi_E^{\parallel}=0.23 \pm 0.21, 
\end{aligned}
\end{align}
with $\chi^2/\mathrm{d.o.f.}=8.43$. These obtained nonfactorizable parameters
are highly precise, with the exception of the strong phase $\phi_E^{\parallel}$. 
This large uncertainty arises because the $E$ amplitude always appears
alongside the $T$ or $C$ amplitude, and the transverse polarization 
branching fraction is smaller than that of longitude polarization. 
Consequently, no sufficiently precise data involving 
the $E$ amplitude exist to constrain the strong phase $\phi_E^{\parallel}$.

At the factorization scale $\mu=465.14 \, \, \mathrm{MeV}$, the Wilson coefficient 
is $a_{1}=-0.47$ in the $T$ amplitude, which corresponds to the factorizable 
contribution of the $T$ diagram. We find the hierarchies $|C^{0}|> |T^{0}| > |E^{0}|$ 
and $|T^{\parallel(\perp)}|\sim |C^{\parallel(\perp)}| > |E^{\parallel(\perp)}|$.
This indicates that the nonfactorizable contribution of the $C$ diagram is 
comparable to the factorizable contributions of the $T$ diagram. This suggests
the potential need to include nonfactorizable contributions for the $T$ diagram 
by inducing five additional parameters $\chi_T^{0, \parallel, \perp}$ and strong 
phases $\phi_T^{\parallel, \perp}$. However, obtaining reliable results for such 
a large number of parameters is difficult given the limited experimental data.

Using the specific decay modes $D^0 \to  \bar K^{*-} \rho^+$ and
$D^0 \to \rho^0 \bar K^{*0}$ as examples, we examine the hierarchies of 
polarization components ($0,\parallel,$ and $ \perp$) and partial-wave contributions 
($S,P,$ and $D$) in the $T$, $C$ and $E$ amplitudes:
\begin{align}\label{hierarchy1}
\begin{aligned}
|T^{0}_{D^0 \to  K^{*-} \rho^+}|: |T^{\parallel}_{D^0 \to   K^{*-} \rho^+}| : 
|T^{\perp}_{D^0 \to   K^{*-} \rho^+}|&=1:0.97:0.50\, ,\\
|C^{0}_{D^0 \to \rho^0 \, \bar K^{*0}}|: |C^{\parallel}_{D^0 \to \rho^0\,  \bar K^{*0}}| : 
|C^{\perp}_{D^0 \to \rho^0\,  \bar K^{*0}}|&=1:0.81:0.47\, ,\\
|E^{0}_{D^0 \to  K^{*-} \rho^+}|: |E^{\parallel}_{D^0 \to  K^{*-} \rho^+}| &=1:1.31\, ,
\end{aligned}
\end{align}
and
\begin{align}\label{hierarchy2}
\begin{aligned}
|T^{S}_{D^0 \to  K^{*-} \rho^+}|: |T^{P}_{D^0 \to  K^{*-} \rho^+}| : 
|T^{D}_{D^0 \to  K^{*-} \rho^+}|&=1:0.37:0.19\, ,\\
|C^{S}_{D^0 \to \rho^0 \, \bar K^{*0}}|: |C^{P}_{D^0 \to \rho^0\,  \bar K^{*0}}| : 
|C^{D}_{D^0 \to \rho^0\,  \bar K^{*0}}|&=1:0.37:0.28\, ,\\
|E^{S}_{D^0 \to  K^{*-} \rho^+}| : |E^{D}_{D^0 \to  K^{*-} \rho^+}|&=1:1.36\, ,
\end{aligned}
\end{align}
respectively.

The hierarchies in Eq.(\ref{hierarchy1}) indicate that the polarization 
amplitudes satisfy 
$ |\mathcal{A}^0| >  |\mathcal{A}^\parallel |>  |\mathcal{A}^\perp|$ for
the $T$ amplitude and, particularly, for the $C$ amplitude. This differs 
from the naive factorization prediction,
$ |\mathcal{A}^\parallel |  \geq  |\mathcal{A}^0|>  |\mathcal{A}^\perp|$. 
This discrepancy arises because the 
$C$ diagram contribution is primarily nonfactorizable, which cannot 
be accurately calculated within the naive factorization approach.
Our result implies that the longitudinal polarization $f_L$ remains 
dominant in decay modes not involving the $E$ amplitude. For 
decays involving interference between $T$ and $E$, or $C$ and 
$E$ amplitudes, the strong phases $\phi^{0,\parallel}_{E,C}$ 
determine the interference pattern, potentially resulting in 
$f_L < f_\parallel$. 

Regarding partial-wave amplitudes, Eq.(\ref{hierarchy2}) shows that 
$|S|>|P>|D|$ for the $T$ and $C$ diagrams. Thus, the $S$-wave 
contribution is dominant. This is attributed to the definition in 
Eq.(\ref{SPDdef}), where $\mathcal{A}^0$ and $\mathcal{A}^\parallel$ 
exhibit constructive interference for the $S$-wave amplitude but destructive 
interference for the $D$ wave. However, the hierarchy is reversed for the 
$E$ diagram (e.g., $|S|<|D|$) due to the large strong phase difference 
$|\phi_E^0-\phi_E^\parallel|$. The hierarchy of $|S|<|D|$ for $E$ amplitudes 
can lead to a larger $D$-wave branching fraction than that of the $S$ wave in
the decays involving the $E$ diagram, contrary to the naive factorization 
prediction of $S$-wave dominance. 

\subsection{Branching fractions of $D \to V\, V$ decays}
Our numerical predictions for the branching fractions of 28 
decay modes of $D$ meson are collected in 
Tables~\ref{CF}-\ref{DCS1}, organized by decay type: 
Cabibbo-favored (CF,  Table~\ref{CF}), singly 
Cabibbo-suppressed (SCS, Tables~\ref{SCS1}-\ref{SCS2}), 
and doubly Cabibbo-suppressed (DCS, Table \ref{DCS1}). 
In each table, the first column lists the decay modes, followed 
by the total branching fraction and its partial-wave components 
($S$ waves, $P$ waves, and $D$ waves). The second column 
identifies the relevant topological diagram amplitudes ($T, C, E$, 
and $A$) for the convenience of the subsequent discussion.
The theoretical results, denoted as ``$\mathcal{B}_{\text{FAT}}$",
appear in the final column. The quoted uncertainties correspond to
(i) variations in the nonperturbative parameters of Eq.~(\ref{parameter});
(ii) a  $10\%$ variation in the form factors; and 
(iii) a  $5\%$ variation in the unmeasured decay constants.
Among these, the form factor uncertainty dominates. For comparison, 
we also list the experimental data (``$\mathcal{B}_{\text{exp}}$" ), 
which constitute the dataset used in our global fit. We explicitly exclude
the $D^0 \to \bar{K}^{*0} K^{*0}$ channel from the fit. Including this 
mode would result in a $\chi^2$ contribution of approximately 484 
from this single mode alone, significantly compromising the global fit 
quality. This discrepancy arises because, at the quark level, the decay 
amplitude is governed by the $E$ diagram with a CKM structure of 
$V_{cd}^*V_{ud}\, +V_{cs}^* V_{us}\, $. Because of the unitarity of the 
CKM matrix, these two terms undergo significant cancellation, 
leading to a theoretically suppressed branching fraction that falls 
well short of the the experimental measurement 
$\mathcal{B} (D^0 \to \bar{K}^{*0} K^{*0})=
(0.88\pm0.04) \times 10^{-3}$~\cite{ParticleDataGroup:2024cfk}.
With more precise data becoming available in the future, 
we will be able to incorporate additional SU(3)-breaking effects.
For instance, by introducing two distinct sets of 
parameters for $E$ amplitude (e.g. $\chi_q^E,\, \chi_s^E$ and $\phi_q^E,\, \phi_s^E$),
we can differentiate between strongly produced light-quark pair 
($u\,  \bar u$ or $d\, \bar d $) or strange quarks ($s\, \bar s$), as applied to
$D \to PP$~\cite{Li:2012cfa} and $D \to PV$  \cite{Li:2013xsa, Zheng:2025ryf}. 
Consequently, the amplitude
$E^{i} \propto V_{cd}^*V_{ud}\, \chi_s^{i}  \phi_s^{i} +V_{cs}^* V_{us}\,\chi_q^{i}  \phi_q^{i} $
(with $i= 0,\parallel, \perp$) will not longer be canceled due to the differences between 
$\chi_q^E,\, \chi_s^E$ and $\phi_q^E,\, \phi_s^E$, potentially explaining the current 
experimental data.
\begin{table}[htbp]
\centering
\caption{\label{CF} Total and partial-waves ($S,P$, and $D$) branching fractions
for Cabibbo-favored $D \rightarrow V V$ decays, given in units of percentage. 
Our theoretical predictions, denoted as $\mathcal{B}_\mathrm{FAT}$, including
uncertainties arising from the fitted parameters, form factors and decay constants, 
respectively, are compared with experimental data, $\mathcal{B}_{\text{exp}}$.
The second column indicates the topological diagram amplitudes, $T, C, E$ , and $A$, 
contributing to each mode. }
\vspace{0.15 cm}
\scalebox{0.98}{
\begin{tabular}{lcccc}
\hline \hline
~~~~Modes ~~~~& ~~~~Amplitudes ~~~~&~~~~$\mathcal{B}_{\text{exp}}$~~~~&~~~~$\mathcal{B}_{\text{FAT}}$~~~~  \\
\hline
$D \rightarrow V_1 V_2$ &$V_{cs}^* V_{ud}$&  \\
$D^0 \rightarrow{K^{*-} \rho^+}$ & $T+E$ &~~~~$6.5\pm 2.5$~~~~&~~~~$4.23\pm0.82\pm0.50\pm0.28$  \\
$D^0 \rightarrow{K^{*-} \rho^+}(S)$ &  &~~~~$1.4\pm 0.4$~~~~&~~~~$2.07\pm0.51\pm0.36\pm0.30$  \\
$D^0 \rightarrow{K^{*-} \rho^+}(P)$ &&~~~~$0.9\pm 0.2$~~~~&~~~~$0.46\pm0.02\pm0.09\pm0.05$  \\
$D^0 \rightarrow{K^{*-} \rho^+}(D)$ &  &~~~~$2.9\pm 0.8$~~~~&~~~~$1.70\pm0.50\pm0.34\pm0.20$ \\
\hline
$D^0 \rightarrow {\rho^0 \bar{K}^{*0} }$& $\frac{1}{\sqrt{2}}(C-E)$&~~~~$1.52\pm0.08$~~~~&~~~~$1.65\pm0.40\pm0.22\pm0.17$ \\
$D^0 \rightarrow { \rho^0\bar{K}^{*0}}(S)$& &~~~~$0.8\pm0.12 $~~~~&~~~~$0.68\pm0.23\pm0.16\pm0.09$  \\
$D^0 \rightarrow {\rho^0 \bar{K}^{*0} }(P)$&&~~~~$0.28\pm0.03$~~~~&~~~~$0.25\pm0.02\pm0.05\pm0.03$  \\
$D^0 \rightarrow {\rho^0 \bar{K}^{*0} }(D)$&  &~~~~$0.98\pm0.1$~~~~&~~~~$0.72\pm0.27\pm0.12\pm0.10$ \\
\hline
$D^0 \rightarrow {\omega \bar{K}^{*0} }$&$\frac{1}{\sqrt{2}}(C+E)$&~~~~$1.1\pm0.5$~~~~ &~~~~$3.63\pm0.43\pm0.34\pm0.44$ \\
$D^0 \rightarrow {\omega \bar{K}^{*0} }(S)$&&&~~~~$3.01\pm0.40\pm0.33\pm0.35$ \\
$D^0 \rightarrow {\omega \bar{K}^{*0} }(P)$&&&~~~~$0.25\pm0.02\pm0.05\pm0.02$ \\
$D^0 \rightarrow {\omega \bar{K}^{*0} }(D)$&&&~~~~$0.37\pm0.18\pm0.04\pm0.08$ \\
\hline

$D^+ \rightarrow {\bar{K}^{*0} \rho^+}$& $T+C$ &  ~~~~$6.23\pm0.27$~~~~&~~~~$6.32\pm1.05\pm0.66\pm0.45$ \\
$D^+ \rightarrow {\bar{K}^{*0} \rho^+}(S)$&&~~~~$6.05\pm0.30$~~~~&~~~~$5.61\pm0.99\pm0.57\pm0.40$  \\
$D^+ \rightarrow {\bar{K}^{*0} \rho^+}(P)$&  &  ~~~~$0.173\pm0.026$~~~~&~~~~$0.16\pm0.03\pm0.03\pm0.01$\\
$D^+ \rightarrow {\bar{K}^{*0} \rho^+}(D)$&  &  &~~~~$0.55\pm0.29\pm0.23\pm0.05$ \\
\hline
$D^+_s \rightarrow{\phi \,\ \rho^+}$& $T$  &  ~~~~$3.98\pm0.39$~~~~&~~~~$4.55\pm0.19\pm0.58\pm0.46$ \\
$D^+_s \rightarrow{\phi \,\ \rho^+}(S)$&& ~~~~ $3.32\pm0.35$~~~~&~~~~$3.85\pm0.16\pm0.55\pm0.38$\\
$D^+_s \rightarrow{\phi \,\ \rho^+}(P)$& &  ~~~~$0.63\pm0.13$~~~~&~~~~$0.53\pm0.02\pm0.11\pm0.05$ \\
$D^+_s \rightarrow{\phi \,\ \rho^+}(D)$& &  &~~~~$0.17\pm0.01\pm0.12\pm0.02$  \\
\hline
$D^+_s \rightarrow {{K}^{*+} \bar{K}^{*0}}$& $C+A$ &  ~~~~$5.93\pm0.88$~~~~&~~~~$5.07\pm0.36\pm0.67\pm0.51$ \\
$D^+_s \rightarrow {{K}^{*+} \bar{K}^{*0}}(S)$&& ~~~~ $5.01\pm0.92$~~~~&~~~~$4.12\pm0.34\pm0.58\pm0.41$ \\
$D^+_s \rightarrow {{K}^{*+} \bar{K}^{*0}}(P)$& & ~~~~ $1.10\pm0.19$~~~~&~~~~$0.56\pm0.06\pm0.11\pm0.06$ \\
$D^+_s \rightarrow {{K}^{*+} \bar{K}^{*0}}(D)$& &  ~~~~$0.65\pm0.16$~~~~&~~~~$0.40\pm0.11\pm0.20\pm0.04 $\\
\hline \hline
\end{tabular}
}
\end{table}
\begin{table}[htbp]
\centering
\caption{\label{SCS1} 
The same as Table \ref{CF} for singly Cabibbo-suppressed $D^0 \rightarrow V V$ decays in units of $10^{-3}$.}
\vspace{0.15 cm}
\scalebox{0.98}{
\begin{tabular}{lccccc}
\hline \hline
~~~~Modes ~~~~& ~~~~Amplitudes ~~~~&~~~~$\mathcal{B}_{\text{exp}}$~~~~&~~~~$\mathcal{B}_{\text{FAT}}$~~~~  \\
\hline
$D \rightarrow V_1 V_2$ &$V_{cd}^*V_{ud}$ or $V_{cs}^* V_{us}$&   &\\
$D^0 \rightarrow { \rho^- \rho^+}$& $T+E$&~~~~$7.81\pm 1.14$~~~~&~~~~$2.67\pm0.54\pm0.30\pm0.21$\\
$D^0 \rightarrow { \rho^- \rho^+}(S)$&&~~~~$1.20\pm 0.40$~~~~&~~~~$1.08\pm0.31\pm0.18\pm0.13$ \\
$D^0 \rightarrow { \rho^- \rho^+}(P)$& &~~~~$1.80\pm 0.30$~~~~&~~~~$0.39\pm0.02\pm0.08\pm0.04$\\
$D^0 \rightarrow { \rho^- \rho^+}(D)$& &~~~~$3.30\pm 0.50$~~~~&~~~~$1.20\pm0.35\pm0.21\pm0.09$\\
\hline

$D^0 \rightarrow { \rho^0 \rho^0}$& $\frac{1}{2}(E-C)$&~~~~$1.33\pm 0.35$~~~~&~~~~$0.49\pm0.13\pm0.07\pm0.04$\\
$D^0 \rightarrow { \rho^0 \rho^0}(S)$& &~~~~$0.18\pm 0.13$~~~~&~~~~$0.18\pm0.07\pm0.04\pm0.03$\\
$D^0 \rightarrow { \rho^0 \rho^0}(P)$&  &~~~~$0.53\pm 0.13$~~~~ &~~~~$0.061\pm0.006\pm0.012\pm0.006$\\
$D^0 \rightarrow { \rho^0 \rho^0}(D)$&&~~~~$0.62\pm 0.3$~~~~&~~~~$0.26\pm0.09\pm0.04\pm0.02$\\
\hline
$D^0 \rightarrow {{K}^{*-} {K}^{*+}}$&$T+E$ & &~~~~$1.58\pm0.28\pm0.19\pm0.19$&\\
$D^0 \rightarrow {{K}^{*-} {K}^{*+}}(S)$&& &~~~~$0.85\pm0.19\pm0.15\pm0.08$\\
$D^0 \rightarrow {{K}^{*-} {K}^{*+}}(P)$& & &~~~~$0.22\pm0.01\pm0.04\pm0.02$\\
$D^0 \rightarrow {{K}^{*-} {K}^{*+}}(D)$& & &~~~~$0.52\pm0.16\pm0.12\pm0.13$\\
\hline

$D^0 \rightarrow { \rho^0 \phi}$& $\frac{1}{\sqrt{2}}C$ &~~~~$1.56\pm 0.13$~~~~ &~~~~$1.13\pm0.08\pm0.14\pm0.11$\\
$D^0 \rightarrow { \rho^0 \phi}(S)$&&~~~~$1.40\pm 0.10$~~~~ &~~~~$0.92\pm0.08\pm0.13\pm0.09$\\
$D^0 \rightarrow { \rho^0 \phi}(P)$&  &~~~~$0.081\pm 0.039$~~~~&~~~~$0.149\pm0.015\pm0.030\pm0.015$\\
$D^0 \rightarrow { \rho^0 \phi}(D)$&  &~~~~$0.085\pm 0.028$~~~~ &~~~~$0.053\pm 0.019\pm0.032\pm0.005$\\
\hline

$D^0 \rightarrow {\omega \phi}$& $\frac{1}{\sqrt{2}}C$ &~~~~$0.648\pm 0.104$~~~~ &~~~~$1.066\pm0.074\pm0.135\pm0.107$\\
$D^0 \rightarrow {\omega \phi}(S)$& &&~~~~$0.873\pm 0.073\pm0.124\pm0.087$ \\
$D^0 \rightarrow {\omega \phi}(P)$& &&~~~~$0.141\pm 0.014\pm0.028\pm0.014$ \\
$D^0 \rightarrow {\omega \phi}(D)$& &&~~~~$0.051\pm 0.018\pm0.031\pm0.005$ \\
\hline

$D^0 \rightarrow {\omega \rho^{0}}$& $-E$&  &~~~~$0.77\pm0.19\pm 0.08\pm 0.08$\\
$D^0 \rightarrow {\omega \rho^{0}}(S)$&&  &~~~~$0.35\pm 0.14\pm 0.08\pm0.03$\\
$D^0 \rightarrow {\omega \rho^{0}}(P)$& &  &~~~~$0.0019\pm 0.0002\pm 0.0027\pm0.0014$\\
$D^0 \rightarrow {\omega \rho^{0}}(D)$& &  &~~~~$0.42\pm 0.17\pm0.02\pm 0.07$\\
\hline

$D^0 \rightarrow { \omega \omega}$& $\frac{1}{2}(C+E)$&  & ~~~~$0.71\pm 0.09\pm0.07\pm 0.10$\\
$D^0 \rightarrow { \omega \omega}(S)$&&  & ~~~~$0.58\pm0.08\pm0.06\pm0.08$\\
$D^0 \rightarrow { \omega \omega}(P)$& &  &~~~~ $ 0.041\pm 0.004\pm 0.008 \pm0.004$\\
$D^0 \rightarrow { \omega \omega}(D)$& &  & ~~~~$0.09\pm0.041\pm0.01 \pm0.02$&\\
\hline \hline
\end{tabular}
}
\end{table}

\begin{table}[htbp]
\centering
\caption{\label{SCS2}The same as Table \ref{CF} for singly Cabibbo-suppressed $D_{(s)}^+ \rightarrow V V$ decays in units of $10^{-3}$. }
\vspace{0.15 cm}
\scalebox{0.87}{
\begin{tabular}{lccccc}
\hline \hline
~~~~Modes ~~~~& ~~~~Amplitudes ~~~~&~~~~$\mathcal{B}_{\text{exp}}$~~~~&~~~~$\mathcal{B}_{\text{FAT}}$~~~~  \\
\hline
$D \rightarrow V_1 V_2$ &$V_{cd}^*V_{ud}$ or $V_{cs}^* V_{us}$&   &\\
$D^+ \rightarrow {\rho^0 \rho^+}$& $-\frac{1}{\sqrt{2}}(T+C)$& &~~~~$ 1.81\pm0.30\pm 0.26\pm0.18$\\
$D^+ \rightarrow {\rho^0 \rho^+}(S)$&& &~~~~$1.53\pm0.27\pm 0.22\pm 0.15$\\
$D^+ \rightarrow {\rho^0 \rho^+}(P)$& & &~~~~$0.07\pm 0.01\pm0.01\pm 0.01$\\
$D^+ \rightarrow {\rho^0 \rho^+}(D)$& & &~~~~$0.21\pm 0.10\pm0.09\pm 0.02$&\\
\hline
$D^+ \rightarrow {\bar{K}^{*0} {K}^{*+}}$& $T$& &~~~~$4.22\pm0.18\pm0.53\pm0.42$\\
$D^+ \rightarrow {\bar{K}^{*0} {K}^{*+}}(S)$&& &~~~~$3.61\pm0.15\pm0.52\pm0.36$\\
$D^+ \rightarrow {\bar{K}^{*0} {K}^{*+}}(P)$& & &~~~~$0.54\pm0.02\pm0.11\pm0.05$\\
$D^+ \rightarrow {\bar{K}^{*0} {K}^{*+}}(D)$& & &~~~~$0.060\pm0.003\pm0.065\pm0.006$\\
\hline
$D^+ \rightarrow {\omega \rho^+}$&$\frac{1}{\sqrt{2}}(T+C+2A)$ & &~~~~$1.52\pm 0.24\pm 0.16\pm 0.12$\\
$D^+ \rightarrow {\omega \rho^+}(S)$&& &~~~~$1.27\pm 0.21\pm 0.14\pm 0.10$\\
$D^+ \rightarrow {\omega \rho^+}(P)$& & &~~~~$0.10\pm 0.02\pm 0.04 \pm 0.02$\\
$D^+ \rightarrow {\omega \rho^+}(D)$& & &~~~~$0.15\pm 0.07\pm 0.05 \pm 0.01$\\
\hline
$D^+ \rightarrow { \rho^+ \phi}$& $C$&  &~~~~$5.88\pm0.41\pm0.74\pm0.59$\\
$D^+ \rightarrow { \rho^+ \phi}(S)$&&  &~~~~$4.82\pm0.40\pm0.69\pm0.48$\\
$D^+ \rightarrow { \rho^+ \phi}(P)$& &  &~~~~$0.78\pm0.08\pm0.16\pm0.08$\\
$D^+ \rightarrow { \rho^+ \phi}(D)$& &  &~~~~$0.28\pm0.10\pm0.17\pm0.03$\\
\hline

$D^+_s \rightarrow { {K}^{*0} \rho^+}$&$T+A$  &~~~~$3.95 \pm 0.39$~~~~&~~~~$2.63\pm0.11\pm0.33\pm0.26$\\
$D^+_s \rightarrow { {K}^{*0} \rho^+}(S)$&&~~~~$1.41 \pm 0.24$~~~~&~~~~$2.06\pm0.09\pm0.29\pm0.21$\\
$D^+_s \rightarrow { {K}^{*0} \rho^+}(P)$&  &~~~~$2.53 \pm 0.31$~~~~ &~~~~$0.43\pm0.02\pm0.09\pm0.04$\\
$D^+_s \rightarrow { {K}^{*0} \rho^+}(D)$&  & &~~~~$0.138\pm0.006\pm0.080\pm0.014$ \\
\hline
$D^+_s \rightarrow { {K}^{*+} \rho^0}$& $\frac{1}{\sqrt{2}}(A-C)$&&~~~~$1.46\pm0.11\pm0.20\pm0.15$ \\
$D^+_s \rightarrow { {K}^{*+} \rho^0}(S)$& &&~~~~$1.17\pm0.09\pm0.17\pm0.12$ \\
$D^+_s \rightarrow { {K}^{*+} \rho^0}(P)$& &~~~~$0.42\pm 0.17$~~~~&~~~~$0.14\pm0.01\pm0.03\pm0.01$  \\
$D^+_s \rightarrow { {K}^{*+} \rho^0}(D)$& &&~~~~$0.16\pm0.04\pm0.07\pm0.02$ \\
\hline
$D^+_s \rightarrow { {K}^{*+} \omega}$&$\frac{1}{\sqrt{2}}(C+A)$ & &~~~~$0.98\pm0.07\pm0.13\pm0.10$\\
$D^+_s \rightarrow { {K}^{*+} \omega}(S)$&& &~~~~$0.78\pm 0.06\pm0.11\pm0.08$\\
$D^+_s \rightarrow { {K}^{*+} \omega}(P)$& & &~~~~$0.09\pm0.01\pm0.02\pm0.01$\\
$D^+_s \rightarrow { {K}^{*+} \omega}(D)$& & &~~~~$0.10\pm0.03 \pm0.05\pm0.01$\\
\hline
$D^+_s \rightarrow {\phi {K}^{*+}}$& $T+C+A$ & &~~~~$1.26 \pm0.20 \pm0.15 \pm 0.10$\\
$D^+_s \rightarrow {\phi {K}^{*+}}(S)$&& &~~~~$1.11\pm0.19\pm0.13 \pm 0.09$\\
$D^+_s \rightarrow {\phi {K}^{*+}}(P)$& & &~~~~$0.04\pm 0.01\pm 0.01\pm0.01$\\
$D^+_s \rightarrow {\phi {K}^{*+}}(D)$& & &~~~~$0.10\pm0.05\pm0.05\pm0.01$\\
\hline \hline
\end{tabular}
}
\end{table}

\begin{table}[htbp]
\centering
\caption{\label{DCS1}Same as Table \ref{CF} for doubly Cabibbo-suppressed $D \rightarrow V V$ decays in units of $10^{-4}$.  }
\vspace{0.15 cm}
\scalebox{0.98}{
\begin{tabular}{lcccccc}
\hline
~~~~Modes ~~~~& ~~~~Amplitudes ~~~~&~~~~$\mathcal{B}_{\text{FAT}}$~~~~  \\
\hline
$D \rightarrow V_1 V_2$ &$V_{cd}^*V_{us}$&  &\\
$D^0 \rightarrow {\rho^- {K}^{*+}}$& $T+E$& $1.24\pm0.23\pm0.14\pm0.13$\\
$D^0 \rightarrow {\rho^- {K}^{*+}}(S)$&& $0.55\pm0.14\pm0.09\pm0.06$\\
$D^0 \rightarrow {\rho^- {K}^{*+}}(P)$& & $0.23\pm0.01\pm0.05\pm0.02$\\
$D^0 \rightarrow {\rho^- {K}^{*+}}(D)$& & $0.46\pm0.14\pm0.09\pm0.07$\\
\hline
$D^0 \rightarrow {{K}^{*0} \rho^0}$&$\frac{1}{\sqrt{2}}(C-E)$ &  $0.47\pm0.11\pm0.06\pm0.05$\\
$D^0 \rightarrow {{K}^{*0} \rho^0}(S)$& &$0.19\pm0.07\pm0.04\pm0.03$\\
$D^0 \rightarrow {{K}^{*0} \rho^0}(P)$& &  $0.07\pm0.01\pm0.01\pm0.01$\\
$D^0 \rightarrow {{K}^{*0} \rho^0}(D)$& &  $0.20\pm0.08\pm0.04\pm0.03$\\
\hline
$D^0 \rightarrow {{K}^{*0} \omega}$&$\frac{1}{\sqrt{2}}(C+E)$ & $1.03\pm0.12\pm0.10\pm0.12$\\
$D^0 \rightarrow {{K}^{*0} \omega}(S)$&& $0.86\pm0.12\pm0.09\pm0.10$\\
$D^0 \rightarrow {{K}^{*0} \omega}(P)$& & $0.07\pm0.01\pm0.01\pm0.01$\\
$D^0 \rightarrow {{K}^{*0} \omega}(D)$& & $0.11\pm0.05\pm0.01\pm0.02$\\
\hline
$D^+ \rightarrow { {K}^{*0} \rho^+}$& $C+A$&  $3.17\pm0.22\pm0.41\pm0.32$\\
$D^+ \rightarrow { {K}^{*0} \rho^+}(S)$&&  $2.59\pm0.21\pm0.37\pm0.26$\\
$D^+ \rightarrow { {K}^{*0} \rho^+}(P)$& &  $0.37\pm0.04\pm0.07\pm0.04$\\
$D^+ \rightarrow { {K}^{*0} \rho^+}(D)$& &  $0.21\pm0.06\pm0.11\pm0.02$\\
\hline
$D^+ \rightarrow {\rho^0 {K}^{*+}}$& $\frac{1}{\sqrt{2}}(A-T)$&  $1.50\pm0.06\pm0.18\pm0.15$\\
$D^+ \rightarrow {\rho^0 {K}^{*+}}(S)$&&  $1.17\pm0.05\pm0.17\pm0.12$\\
$D^+ \rightarrow {\rho^0 {K}^{*+}}(P)$& &  $0.29\pm0.01\pm0.06\pm0.03$\\
$D^+ \rightarrow {\rho^0 {K}^{*+}}(D)$& &  $0.037\pm0.002\pm0.030\pm0.004$\\
\hline
$D^+ \rightarrow {\omega {K}^{*+}}$& $\frac{1}{\sqrt{2}}(T+A)$& $1.47\pm0.06\pm0.18\pm0.15$ &\\
$D^+ \rightarrow {\omega {K}^{*+}}(S)$&& $1.15\pm0.05\pm0.16\pm0.11$\\
$D^+ \rightarrow {\omega {K}^{*+}}(P)$& & $0.29\pm0.01\pm0.06\pm0.03$\\
$D^+ \rightarrow {\omega {K}^{*+}}(D)$& & $0.036\pm0.002\pm0.029\pm0.004$\\
\hline
$D^+_s \rightarrow {{K}^{*0} {K}^{*+}}$&$T+C$ & $0.82\pm0.13\pm0.12\pm0.08$\\
$D^+_s \rightarrow {{K}^{*0} {K}^{*+}}(S)$&& $0.70\pm0.12\pm0.10\pm0.07$\\
$D^+_s \rightarrow {{K}^{*0} {K}^{*+}}(P)$& & $0.037\pm0.007\pm0.007\pm0.004$\\
$D^+_s \rightarrow {{K}^{*0} {K}^{*+}}(D)$& & $0.08\pm0.04\pm0.04\pm0.01$\\
\hline
 \hline
\end{tabular}
}
\end{table} 
A comparison of the values in the last two columns indicates that our 
results are consistent with the measured CF and SCS modes within 
uncertainties, with the exception of the $P$-wave branching fractions 
for $D_s^+ \to K^{*0} \rho^+$ and $D^0 \to \rho^- \rho^+$ decays. 
These two modes contribute significantly to the total $\chi^2$, yielding 
values of approximately 45 for $D_s^+ \to K^{*0} \rho^+$ and 22 for 
$D^0 \to \rho^- \rho^+$. Since the $E^\perp$ is not included in our 
analysis, these two modes receive contributions only from the $T$ 
diagram, corresponding to predictions made using naive factorization.
Additionally, pure annihilation processes, $D_s^+ \to \rho^+ \omega$, 
$D^+ \to \bar{K}^{*0} K^{*+}$, and $D_s^+ \to K^{*+} \rho^0$, mediated 
exclusively by the $A$ diagram are excluded from our predictions due to 
the neglect of $A$ diagram contribution. Kinematically forbidden decays, 
$D^0 \to \bar{K}^{*0} \phi, {K}^{*0} \phi, \phi \phi$ and 
$D^+ \to {K}^{*+} \phi$, are also omitted. Predictions for unmeasured 
decay modes in the FAT approach remain to be tested against future 
experimental data.

In the following, we analyze the branching fraction values
of specific modes listed in Tables~\ref{CF}-\ref{DCS1}. 
The total branching fractions of these modes are dominated 
by contributions from the $T$ and $C$ diagrams. 
The partial-wave results are discussed explicitly for each mode.

We begin with modes where the $D$-wave branching fraction 
exceeds the $S$-wave, such as $D^0 \rightarrow {\rho^0 \bar{K}^{*0} }$, 
$D^0 \rightarrow { \rho^0 \rho^0}$, and $D^0 \rightarrow {{K}^{*0} \rho^0}$, 
which involve the amplitudes $\frac{1}{\sqrt{2}}(C-E)$, as well as 
$D^0 \rightarrow {\omega \rho^{0}}$ involving the $E$ amplitude. 
In these decays, the $C$ and the $E$ amplitudes interfere 
destructively for both the $S$-wave and the $D$-wave branching 
fractions. Because the $D$-wave amplitude for the $E$ diagram 
is significantly larger than that for the $C$ diagram, the combined 
$D$-wave term $\frac{1}{\sqrt{2}}(C-E)$ approximates 
$-\frac{1}{\sqrt{2}}E$. Conversely, strong destructive interference 
in the $S$-wave amplitude suppresses its branching fraction, 
which is slightly smaller than that of the $D$ wave. It is obvious 
that $D^0 \rightarrow {\omega \rho^{0}}$, dominated by the $E$ 
amplitude, follows the hierarchy $|S|<|D|$, yielding a larger 
$D$-wave branching fraction.

 The situation is reversed for the modes $D^0 \rightarrow {\omega \bar{K}^{*0}}$,
 $D^0 \rightarrow { \omega \omega}$, and $D^0 \rightarrow {{K}^{*0} \omega}$ 
 characterized by $\frac{1}{\sqrt{2}}(C+E)$ contributions, where the interference 
 pattern is opposite. Specifically, constructive interference between the $C$ and 
 $E$ amplitudes enhances the $S$-wave branching fractions, whereas destructive 
 interference suppresses the $D$-wave component. These results are consistent 
 with the naive factorization predictions, in which the $S$-wave component is 
 dominant.
 
The third category of decay modes involving the $E$ amplitude 
comprises those governed by $T+E$ amplitudes, including
$D^0 \rightarrow{K^{*-} \rho^+}$, $D^0 \rightarrow { \rho^- \rho^+}$, 
$D^0 \rightarrow {{K}^{*-} {K}^{*+}}$, and $D^0 \rightarrow {\rho^- {K}^{*+}}$.
Here, destructive interference between the $T$ and $E$ amplitudes
in the $S$ wave and constructive interference in the $D$ wave
result in comparable $S$-wave and $D$-wave branching fractions.
Finally, the remaining decays precessing through $T$, $C$ or $T+C$ 
diagrams exhibit a dominant $S$-wave branching fraction.
 
As noted in Ref.~\cite{Cheng:2024hdo}, $D^0 \rightarrow { \rho^0 \phi}$ 
and $D^0 \rightarrow {\omega \phi}$ are expected to have identical 
branching fractions (both total and partial wave) as they share 
the same $C$ topological amplitude. It is necessary to consider 
that final state rescattering (e.g., via $D^0 \to K^{*+} K^{*-}$) 
can contribute differently to these two modes~\cite{Cao:2023csx}. 
In the FAT approach, the nonfactorizable parameters $\chi_C$ 
and $\phi_C$ effectively incorporate such FSIs. However, 
the SU(3)-breaking effects (specifically, 
those from the form factors $A^{D^0 \to \rho}$ and $A^{D^0 \to \omega}$, 
$V^{D^0 \to \rho}$ and $V^{D^0 \to \omega}$)
are insufficient to account for the differing central values of total branching 
fractions for $D^0 \rightarrow { \rho^0 \phi}$ and $D^0 \rightarrow {\omega \phi}$.
Nevertheless, within uncertainties, particularly given the large form factor errors,
the branching fractions predicted by the FAT approach 
for both decays are consistent with experimental measurements.
The $P$ and $D$-wave fractions of $D^0 \rightarrow { \rho^0 \phi}$ 
are also consistent with experimental data,
and align with the FSIs predictions of $P$-wave component in Ref.~\cite{Cao:2023csx}.
The partial-wave components of $D^0 \rightarrow {\omega \phi}$,
however, await precise measurement by future experiments.


\subsection{Polarization fractions of $D \to V\, V$ decays}
\begin{table}[htbp]
\centering
\caption{\label{fL} The polarization fractions of $D \rightarrow V_1 V_2$ decays in units of percentage.  }
\vspace{0.15cm}
\scalebox{1}{
\begin{tabular}{lcccc}
\hline \hline
~~~~Modes ~~~~& ~~~~Amplitudes ~~~~&~~~~$f_L$~~~~&~~~~$f_{\parallel}$~~~~  \\
\hline
$D^0 \rightarrow { \rho^0 \phi}$& $\frac{1}{\sqrt{2}}C$ &~~~~$48.57\pm7.25$~~~~ &~~~~$38.22\pm7.04$\\

$D^0 \rightarrow {\omega \phi}$& $\frac{1}{\sqrt{2}}C$ &~~~~$48.58\pm7.25$~~~~ &~~~~$38.14\pm7.04$\\

$D^+ \rightarrow {\bar{K}^{*0} \rho^+}$& $T+C$ &  ~~~~$60.68\pm9.29$~~~~&~~~~$36.75\pm9.23$ \\

$D^+ \rightarrow {\rho^0 \rho^+}$& $-\frac{1}{\sqrt{2}}(T+C)$&~~~~$64.57\pm9.59$~~~~   &~~~~$31.35\pm9.35$\\

$D^+ \rightarrow {\bar{K}^{*0} {K}^{*+}}$& $T$&~~~~$39.91\pm6.18$~~~~  &~~~~$47.18\pm6.36$\\

$D^+ \rightarrow {\omega \rho^+}$&$\frac{1}{\sqrt{2}}(T+C)$ &~~~~$60.08\pm8.96$~~~~  &~~~~$33.15\pm8.47$\\

$D^+ \rightarrow { \rho^+ \phi}$& $C$& ~~~~$48.71\pm7.25$~~~~  &~~~~$38.07\pm7.03$\\

$D^+ \rightarrow { {K}^{*0} \rho^+}$& $C$& ~~~~ $53.33\pm7.25$~~~~ &~~~~$35.13\pm6.88$\\

$D^+ \rightarrow {\rho^0 {K}^{*+}}$& $\frac{1}{\sqrt{2}}T$&~~~~  $40.65\pm6.02$~~~~ &~~~~$39.69\pm5.98$\\

$D^+ \rightarrow {\omega {K}^{*+}}$& $\frac{1}{\sqrt{2}}T$&~~~~ $40.65\pm6.02$ ~~~~ &~~~~$39.61\pm5.98$\\

$D^+_s \rightarrow{\phi \,\ \rho^+}$& $T$  &  ~~~~$47.67\pm6.42$~~~~&~~~~$40.68\pm6.26$ \\

$D^+_s \rightarrow {{K}^{*+} \bar{K}^{*0}}$& $C$ &  ~~~~$55.70\pm7.19$~~~~&~~~~$33.28\pm6.74$ \\

$D^+_s \rightarrow { {K}^{*0} \rho^+}$&$T$  &~~~~$48.67\pm6.25$~~~~&~~~~$34.82\pm5.78$\\

$D^+_s \rightarrow { {K}^{*+} \rho^0}$& $\frac{1}{\sqrt{2}}C$&~~~~$61.35\pm6.92$~~~~ &~~~~$29.34\pm6.37$ \\

$D^+_s \rightarrow { {K}^{*+} \omega}$&$\frac{1}{\sqrt{2}}C$ &~~~~$60.97\pm6.95$~~~~  &~~~~$29.61\pm6.40$\\

$D^+_s \rightarrow {\phi {K}^{*+}}$& $T+C$ & ~~~~$59.79\pm9.50$~~~~ &~~~~$36.91\pm9.42$\\

$D^+_s \rightarrow {{K}^{*0} {K}^{*+}}$&$T+C$ &~~~~ $61.69\pm9.90$~~~~ &~~~~$33.83\pm9.70$\\
\hline 
$D^0 \rightarrow{K^{*-} \rho^+}$ & $T+E$ &~~~~$84.80\pm5.76$~~~~&~~~~$4.39\pm4.51$  \\

$D^0 \rightarrow { \rho^- \rho^+}$& $T+E$&~~~~$83.19\pm5.22$~~~~&~~~~$2.16\pm2.90$\\

$D^0 \rightarrow {{K}^{*-} {K}^{*+}}$&$T+E$ &~~~~$79.05\pm7.16$~~~~ &~~~~$7.35\pm5.93$\\

$D^0 \rightarrow {\rho^- {K}^{*+}}$& $T+E$&~~~~ $77.69\pm6.33$~~~~ &~~~~$3.75\pm3.82$\\

\hline
$D^0 \rightarrow {\rho^0 \bar{K}^{*0} }$& $\frac{1}{\sqrt{2}}(C-E)$&~~~~$72.23\pm10.23$~~~~&~~~~$12.57\pm8.88$ \\

$D^0 \rightarrow { \rho^0 \rho^0}$& $\frac{1}{2}(E-C)$&~~~~$77.43\pm9.46$~~~~&~~~~$10.32\pm8.13$\\

$D^0 \rightarrow {{K}^{*0} \rho^0}$&$\frac{1}{\sqrt{2}}(C-E)$ &~~~~$72.23\pm10.23$~~~~ &~~~~$12.57\pm8.88$\\

$D^0 \rightarrow {\omega \bar{K}^{*0} }$&$\frac{1}{\sqrt{2}}(C+E)$&~~~~$41.36\pm7.98 $~~~~ &~~~~$51.82\pm7.79$ \\

$D^0 \rightarrow { \omega \omega}$& $\frac{1}{2}(C+E)$& ~~~~$45.43\pm8.32$~~~~  & ~~~~$48.87\pm8.06$\\

$D^0 \rightarrow {{K}^{*0} \omega}$&$\frac{1}{\sqrt{2}}(C+E)$ &~~~~ $41.36\pm7.98$~~~~ &~~~~$51.82\pm7.79$\\

$D^0 \rightarrow {\omega \rho^{0}}$& $E$& ~~~~$38.16\pm14.21$~~~~ &~~~~$61.59\pm14.18$\\
\hline
\hline
\end{tabular}
}
\end{table}
Based on the fitted parameters in Eq.(\ref{parameter}), we 
calculated the polarization fractions $f_L$ and $f_\parallel$ 
for $D \to V\, V$ decays, as listed in Table~\ref{fL}. 
The values for $f_L$ and $f_\parallel$, accompanied by 
their uncertainties summed in quadrature, appear in the 
table's last two columns, respectively. For clarity in the 
following discussion, we categorize the 28 decay modes 
into three groups: those governed exclusively by $T$, $C$, 
or $T+C$ diagrams; those dominated by $T+E$ contributions; 
and those proceeding through $C$ and $E$ diagrams.

According to Eq.(\ref{hierarchy1}), for the first category of 
decay modes mediated by $C$ or $T+C$ diagrams, their 
polarization fractions follow the same hierarchy as the 
nonfactorizable parameters, 
i.e., $\chi_C^{0}>\chi_C^{\parallel}$, which implies 
$f_L > f_\parallel$. In contrast, for modes proceeding
only via the $T$ diagram, the longitudinal polarization 
$f_L$ is comparable to or smaller than $f_\parallel$,
 as observed in the $D^+ \to {\bar{K}^{*0} {K}^{*+}}$ 
decay, based on the naive factorization framework.

For the second category of decay modes, where interference 
occurs between the $T$ and the $E$ amplitudes, their polarization 
fractions are governed by the strong phase differences between 
the corresponding helicity amplitudes, $T^{0,\parallel}$ and 
$E^{0,\parallel}$. According to Eq.(\ref{parameter}), the strong 
phase difference between the $T^0$ and $E^0$ amplitudes is 
less than $\pi/2$, leading to constructive interference that 
enhances $f_L$. Conversely, the strong phase difference between 
the $T^\parallel$ and $E^\parallel$ amplitudes exceeds $\pi/2$,
resulting in destructive interference that suppresses $f_\parallel$.
Consequently, the significant interference between the 
nonfactorizable $E$ diagram and the $T$ diagram drives the 
longitudinal polarization $f_L$ to approximately $80\%$.

Finally, we turn to the third category of modes involving 
interference between the $C$ and $E$ amplitudes. Given 
that the strong phase difference satisfy 
$\phi_E^{0}-\phi_C^{0}> \pi/2$ while 
$\phi_E^{\parallel}-\phi_C^{\parallel}< \pi/2$, we observe 
destructive interference in the longitudinal component and
constructive interference in the transverse component for 
decays with $C+E$ contributions. Consequently, $f_L$ is 
suppressed while $f_\parallel$ is enhanced, yielding 
$f_\parallel >f_L $. This hierarchy also holds for the 
$D^0 \rightarrow {\omega \rho^{0}}$ decay, which is 
governed exclusively by the $E$ amplitude. In contrast,  
for modes dominated by $C-E$ interference, $f_L$ is 
enhanced and $f_\parallel$ is suppressed, resulting in 
a dominant longitudinal polarization of $f_L \sim 70\%$. 
Notably, our prediction for the longitudinal polarization $f_L$ in 
$D^0 \rightarrow { \rho^0 \rho^0}$ decay agrees well with 
the experimental measurement of $(71 \pm 4 \pm 2)\%$
reported by the FOCUS Collaboration~\cite{FOCUS:2007ern}.
In contrast, our result for $D^0 \to K^{*-} K^{*+}$ deviates from 
the experimental value of $f_L=0.468 \pm 0.046 \pm 0.011$~\cite{BESIII:2026dwz} 
by $3.8 \sigma$. This discrepancy is expected to be resolved by 
 incorporating  the nonfactorizable contribution into the $T$ amplitude.

  
\subsection{$D^0-\bar D^0$ mixing parameter $y$ in $D \to V\, V$ decays}

$D^0-\bar D^0$ mixing is conventionally characterized by the parameters 
$x \equiv \Delta m/ \Gamma$ and $y \equiv \Delta \Gamma/ 2 \Gamma$, 
where $\Delta m=m_1-m_2$ and $\Delta \Gamma=\Gamma_1-\Gamma_2$ 
denote the mass and width differences of the mass eigenstates $D_{1,2}$,
respectively, and $\Gamma=(\Gamma_1+\Gamma_2)/2$ is the average width.
Assuming {\it CP} conservation, the mass eigenstates $D_{1,2}$ coincide with 
the {\it CP} eigenstates $| D_{\pm} \rangle =(| D^0 \rangle \pm | \bar D^0 \rangle)/ \sqrt 2$.
Then the parameter $y$ can be formulated as~\cite{Jiang:2017zwr}
\begin{align}\label{Dmixy}
y=\frac{1}{\Gamma}\, \sum_n \eta_\mathrm{CP}(n)\, \rho_n\, \mathcal{R}e
\left[\mathcal{A}(D^0 \to n) \mathcal{A}^*(D^0 \to \bar n) \right]\, ,
\end{align}
or as~\cite{Cheng:2024hdo}
\begin{align}\label{Dmixy2}
y= \sum_n \eta_\mathrm{CKM}(n)\, \eta_\mathrm{CP}(n)\, \mathrm{cos} \delta_n\, 
\sqrt{\mathcal{B}(D^0 \to n) \mathcal{B}(D^0 \to \bar n)}\, ,
\end{align}
where $ \rho_n$ is the phase-space factor for the decay into the final state n. 
For $D \to VV$ decays, the {\it CP} eigenvalue is given by $\eta_{CP}(n)=(-1)^L$,
where $L$ represents the relative orbital angular momentum between the two vector mesons.
The CKM factor is defined as $\eta_\mathrm{CKM}=(-1)^{n_s}$, where $n_s$ denotes
 the number of $s$ and $\bar s$ quarks in the final state.
The parameter $ \delta_n$ corresponds to the strong phase difference 
between the $D^0 \to n$ and $\bar D^0 \to n$ amplitudes.

Using the fitted parameters given in Eq.(\ref{parameter}), we 
calculate the mixing parameter $y$ for $D \to V\, V$ decays via Eq.(\ref{Dmixy2}). 
The results for the individual $S$-, $P$-, and $D$-wave contributions, denoted as $y_{S,P,D}$,
yield
\begin{align}\label{ySPD}
y_{VV,S}=(1.08 \pm 0.09)\%\, ,\quad  y_{VV,P}=(-0.22\pm0.02)\%\, ,
\quad y_{VV,D}=(0.55\pm 0.08)\%\, ,
\end{align}
where the uncertainties are obtained by summing 
those from the nonfactorizable parameters, form factors, and decay constants 
in quadrature. 
Compared to the naive factorization predictions in Ref.~\cite{Cheng:2024hdo},
our results for $y_{VV, S,P}$ and $y_{VV, D}$ in Eq.(\ref{ySPD}) are larger
by one and two orders of magnitude, respectively. Notably, 
the predictions for $y_{VV, S,P,D}$ in Ref.~\cite{Cheng:2024hdo} are 
substantially modified when experimental data for measured modes are included,  
alongside naive factorization predictions for the unmeasured ones.
In this case, their values of $y_{VV,P}=-0.0958\%$ and  $ y_{VV,D}=0.361\%$ 
are in much better agreement with our results.

Regarding the longitudinal component $y_L$,
Ref.~\cite{Jiang:2017zwr} evaluated it using the nonfactorizable parameters $\chi^C_V, \, \phi^C_V$
and $\chi^E_{q(s)}, \, \chi^E_{q(s)}$ fitted to data of $D \to PV$ decays
as an approximation.
 Their result, $ y_{VV,L}=(-0.42 \pm 0.34) \times 10^{-3}$, is significantly 
smaller than our value of $ y_{VV,L}=(1.21\pm0.12)\%$, primarily due to
the different nonfactorizable parameters employed.


 \section{Conclusion}\label{sec:4}
Within the FAT framework, we perform a systematical analysis
of $D \to VV $ decays. Four types of topological diagrams, $T$, $C$, 
$E$, and $A$, contribute to these decays through weak interactions. 
The $T$ amplitude is calculated using naive factorization. However, 
the nonfactorizable $C$ and $E$ amplitudes are described by only 
two sets of universal parameters,
$\chi_C^{0, \, \parallel, \, \perp},\phi_C^{0, \, \parallel, \, \perp} $ 
and $\chi_E^{0, \, \parallel}, \phi_E^{0, \, \parallel}$, respectively,
for all $D \to VV $ modes after form factors and decay constants 
are factored out. This demonstrates that SU(3)-breaking effects 
are naturally incorporated in the FAT approach, which keeps the 
number of free parameters minimal. Specifically, these parameters 
characterizing the nonfactorizable contributions of the $C$ and $E$ 
amplitudes, together with the factorization scale $\mu$ of the 
Wilson coefficient $a_1(\mu)$ for $T$ amplitude, constitute 11 free 
parameters fitted globally to 36 experimental data points. The 
obtained nonfactorizable parameters are determined with high 
precision, except for $\phi_E^{\parallel}$.

The fitted parameters yield topological-amplitude hierarchies of
$|C^{0}|> |T^{0}| > |E^{0}|$ and $|T^{\parallel(\perp)}|\sim 
|C^{\parallel(\perp)}| > |E^{\parallel(\perp)}|$, indicating that the 
nonfactorizable $C$ contribution is comparable to the factorizable 
$T$  contributions. Polarization amplitudes follow $|\mathcal{A}^0| 
>  |\mathcal{A}^\parallel |>  |\mathcal{A}^\perp|$ for the $T$ and 
$C$ amplitudes, contrary to the naive factorization prediction.
Thus, the longitudinal polarization $f_L$ dominates in modes 
governed by the $T$ or $C$ amplitudes. However, in decays 
with $T$ and $E$, or $C$ and $E$ interference, strong phases 
$\phi^{0,\parallel}_{E,C}$ can possibly resulting in $f_L < f_\parallel$. 
Finally, the partial-wave hierarchy $|S|<|D|$ yields a dominant 
$D$-wave branching fraction in $E$ diagram mediated decays,
contrary to the naive $S$-wave dominance prediction. 

With the fitted nonfactorizable parameters, we predict the branching 
fractions and polarization fractions for 28 decay modes of $D \to V\, V$.
Our results are in agreement with the current experimental data within 
uncertainties. For modes involving the $C-E$ amplitudes, such as 
$D^0 \rightarrow {\rho^0 \bar{K}^{*0} }$, and modes governed by $E$, 
like $D^0 \rightarrow {\omega \rho^{0}}$, the $S$-wave branching 
fractions are slightly smaller than those of the $D$ wave. This is 
contrary to naive factorization estimations but agrees with current 
observations. 
The polarization fractions $f_L$ of $D^0 \rightarrow { \rho^0 \rho^0}$
is in well agreement with the experimental data.
Predictions for 
unobserved decay modes await for test by upcoming experimental 
studies, especially those with branching fractions on order of 
$10^{-3}-10^{-2}$, the $D$-wave dominated modes, and modes 
exhibiting $f_\parallel >f_L $.
We also calculate the $D^0-\bar D^0$ mixing parameters $y_{S,P,D}$ and $y_L$,
which are of the order of $10^{-3}-10^{-2}$.

\section*{Acknowledgments}
We are grateful to Chao Wang for useful discussion. The work is supported 
by the National Natural Science Foundation of China under Grants No.12465017,
No.12105148, No.12075126.

\bibliographystyle{bibstyle}
\bibliography{refs}

\end{document}